\documentclass[fleqn,10pt]{wlscirep}

\title{Optical Properties of Drug Metabolites in Latent Fingermarks }

\author[1,*]{Yao Shen}
\author[2]{Qing Ai}

\affil[1]{School of Forensic Science, People's Public Security
University of China, Beijing 100038, China}

\affil[2]{Department of Physics, Applied Optics Beijing Area Major Laboratory,
Beijing Normal University, Beijing 100875, China}

\affil[*]{shenyaophysics@hotmail.com}


\begin{abstract}
Drug metabolites usually have structures of split-ring resonators
(SRRs), which might lead to negative permittivity and permeability
in electromagnetic field. As a result, in the UV-vis region, the latent
fingermarks images of drug addicts and non
drug users are inverse. The optical properties of latent fingermarks
are quite different between drug addicts and non-drug users. This
is a technic superiority for crime scene investigation to distinguish
them. In this paper, we calculate the permittivity and permeability
of drug metabolites using tight-binding model. The latent fingermarks
of smokers and non-smokers are given as an example.
\end{abstract}

\begin{document}

\flushbottom
\maketitle

\thispagestyle{empty}


In 1968, negative index material (NIM) was first introduced by Veselago
\cite{x1}. NIM can have negative permittivity
and permeability simultaneously. Pendry \textit{et al.} \cite{x2,x3,x4,x5,Luo13}
gave a deep discussion and pointed out that a configuration which
was called split-ring resonator (SRR) \cite{x3} could put negative
refraction into practice, apart from some particular configurations
with non-trivial symmetry breaking \cite{Fang15}. From then on, negative
refraction became a focus in scientific research, for example NIM
can be used to fabricate perfect lenses to enhance local field and
detection sensitivity \cite{x6-1,x6-2,x6-3,x6-4,Ma10}. Two years later,
Shelby \textit{et al.} \cite{x6} realized NIM experimentally. Metamaterial,
made up of SRRs or molecules which consist of SRRs, e.g. extended
metal atom chains, becomes a new branch of study \cite{x7,x8,x9,x10,x10-1,x10-2,x10-3,x10-4,x10-5,x10-6,x10-7,x10-10,x10-11,x10-12,x10-13,x10-14,x10-15,me-1,me-2,me-3,Zhang14,sr1,sr2,sr3,sr4}.
Many new directions are developed, such as electromagnetic cloaking
\cite{e1,e2,e2-1,e2-2,e2-3,e2-4}, toroidal moment \cite{e3}, liquid
crystal magnetic control \cite{e4}, etc. On the other hand, in forensic science, on
highly reflective surface, the latent fingermarks are difficult to
be observed. The traditional method of visualizing the invisible fingermarks
is using fluorescent tag. In Boddis and Russell's paper \cite{finger},
they made use of antibody-magnetic particle conjugates to visualize them. During this procedure,
they find the latent fingermarks of smokers and non-smokers are quite different. As a result, the latent
fingermarks of these two kinds of donors are observed to be inverse
and thus they can be used to identify the smokers. The research of distinguishing drug users by metabolites becomes a new focus in forensic science field \cite{sr00,sr01}.  In this  paper, we first introduce negative refraction phenomenon to forensic science. We point out that
when we put those latent fingermarks of drug addicts and non-drug users in the light field,
they can also be identified. Furthermore,
our method is physical and non-damaged, because the latent fingermarks will
not be destroyed. More importantly, due to quantum effect, a small
volume of molecules could sufficiently respond negatively to the applied
electromagnetic fields \cite{Dong15}. We give the theoretical derivation
and calculation of this phenomenon. Our result is not only suitable
for smokers but also for drug addicts. In other words, except for
cotinine, the metabolite of nicotine, benzoylecgonine and morphine
can also be detected using our method.

This paper is organized as following: In section~II, we give the
theoretical derivation of the permittivity and permeability of cotinine
on the basis of tight-binding model. And we discuss the
numerical results of permittivity and permeability. Finally, the main
results are concluded and future work is discussed in the discussion section.

\section*{Results}


\subsection*{Tight-binding Approximation and H\"{u}ckel Model}

Many molecules of drug metabolites have a broken ring configuration,
i.e. SRRs. This structure gives them special optical properties. Without
loss of generality, we calculate cotinine, i.e. the metabolite of
nicotine, as an example.

\begin{figure}
\includegraphics[bb=100bp 350bp 500bp 650bp,clip,scale=0.6]{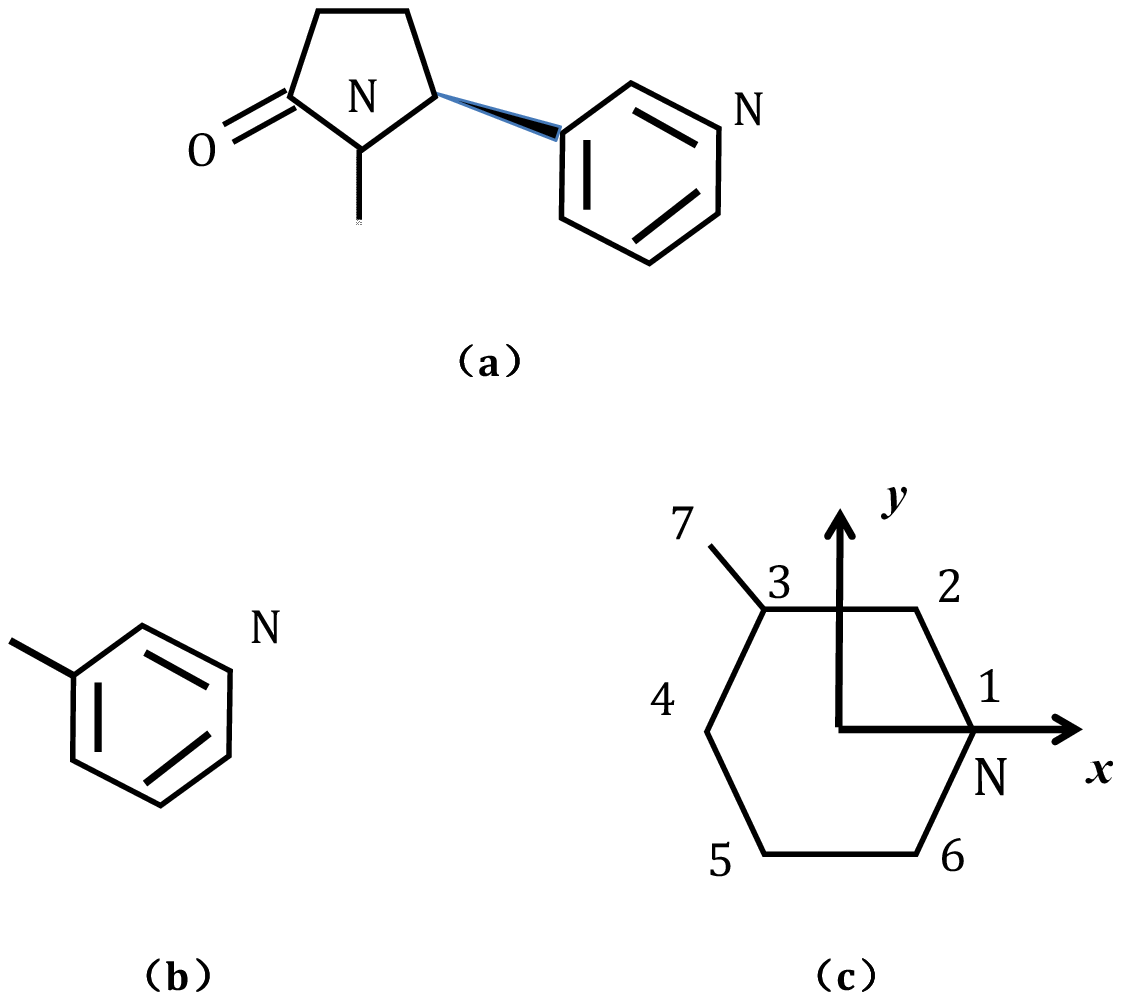}\protect\caption{(a) The chemical structure of cotinine molecule ($C{}_{10}$$H{}_{12}$$N{}_{2}$$O$).
Cotinine has two rings, one is pyridyl, and the other is pyrrolidin.
(b) The simplified model. The molecule is simplified into a pyridyl
and a methyl. (c) The spatial distribution of atoms in the simplified
model. The origin is set at the center of the hexagon. Seven sites
are labeled sequentially. Site $1$ is a nitrogen atom and others
are carbon atoms.\label{fig:two}}
\end{figure}

Figure \ref{fig:two}(a) demonstrate the structure of cotinine molecule.
The main part of cotinine is the hexagon part which is called pyridine
(see Figure \ref{fig:two} (b)). In this part, one carbon atom of
the ring is substituted with one nitrogen atom, and for the sake of
simplicity the remaining part of cotinine molecule is simplified into
a methyl in the same plane. This simplification is reasonable because
the main contribution to the optical property comes from the $\pi$
electrons of conjugate part of cotinine molecule (single-nitrogen-substituted
heterocyclic annulene \cite{x11}).

$\pi$ electrons of the pyridyl interact with the electromagnetic
fields which results in special optical properties of cotinine. The
structure consists of one pyridyl and one methyl with sites labeled
as Figure \ref{fig:two}(c). The quantum dynamics of $\pi$ electrons
around these seven sites are described by the H\"{u}ckel model as
\cite{x12}
\begin{equation}
\mathcal{H}=\sum_{j=1}^{6}\alpha_{j}\left|j\right\rangle \left\langle j\right|+\alpha_{C}\left|7\right\rangle \left\langle 7\right|+\sum_{j=1}^{6}\beta_{j,j+1}\left(\left|j\right\rangle \left\langle j+1\right|+\left|j+1\right\rangle \left\langle j\right|\right)+\beta_{CC}\left(\left|3\right\rangle \left\langle 7\right|+\left|7\right\rangle \left\langle 3\right|\right),\label{eq:Hhu}
\end{equation}
where$j$ is the site label, $\left|j\right\rangle $ denotes the
state with a $\pi$ electron at site $j$ , $\alpha_{j}$ is the site
energy, the coupling strength between $j$th and $(j+1)$th sites
is given by the resonant integral $\beta_{j,j+1}$. Here, when $j=6$,
the adjacent site is not site 7 but site 1. The nitrogen atom is located
at site 1 (see Figure \ref{fig:two}(c)). In this configuration, the
site energies and coupling constants are explicitly given as
\begin{eqnarray}
\alpha_{j} & = & \left\{ \begin{array}{c}
\alpha_{C}\textrm{, for }j\neq1,\\
\alpha_{N}\textrm{, for }j=1,
\end{array}\right.\\
\beta_{j,j+1} & = & \left\{ \begin{array}{c}
\beta_{CC}\textrm{, for }j\neq1,6,\\
\beta_{CN}\textrm{, for }j=1,6.
\end{array}\right.
\end{eqnarray}

By diagonalization, the H\"{u}ckel Hamiltonian~(\ref{eq:Hhu}) can
be reexpressed as
\begin{eqnarray}
\mathcal{H} & = & \sum_{k=1}^{7}\varepsilon_{k}\left|\psi_{k}\right\rangle \left\langle \psi_{k}\right|,
\end{eqnarray}
where
\[
\left|\psi_{k}\right\rangle =\sum_{j=1}^{7}C_{ki}\left|j\right\rangle
\]
is $k$th single-electron molecular orbital, $\varepsilon_{k}$ is
energy level.

\begin{figure}
\includegraphics[bb=50bp 400bp 595bp 782bp,clip,scale=0.6]{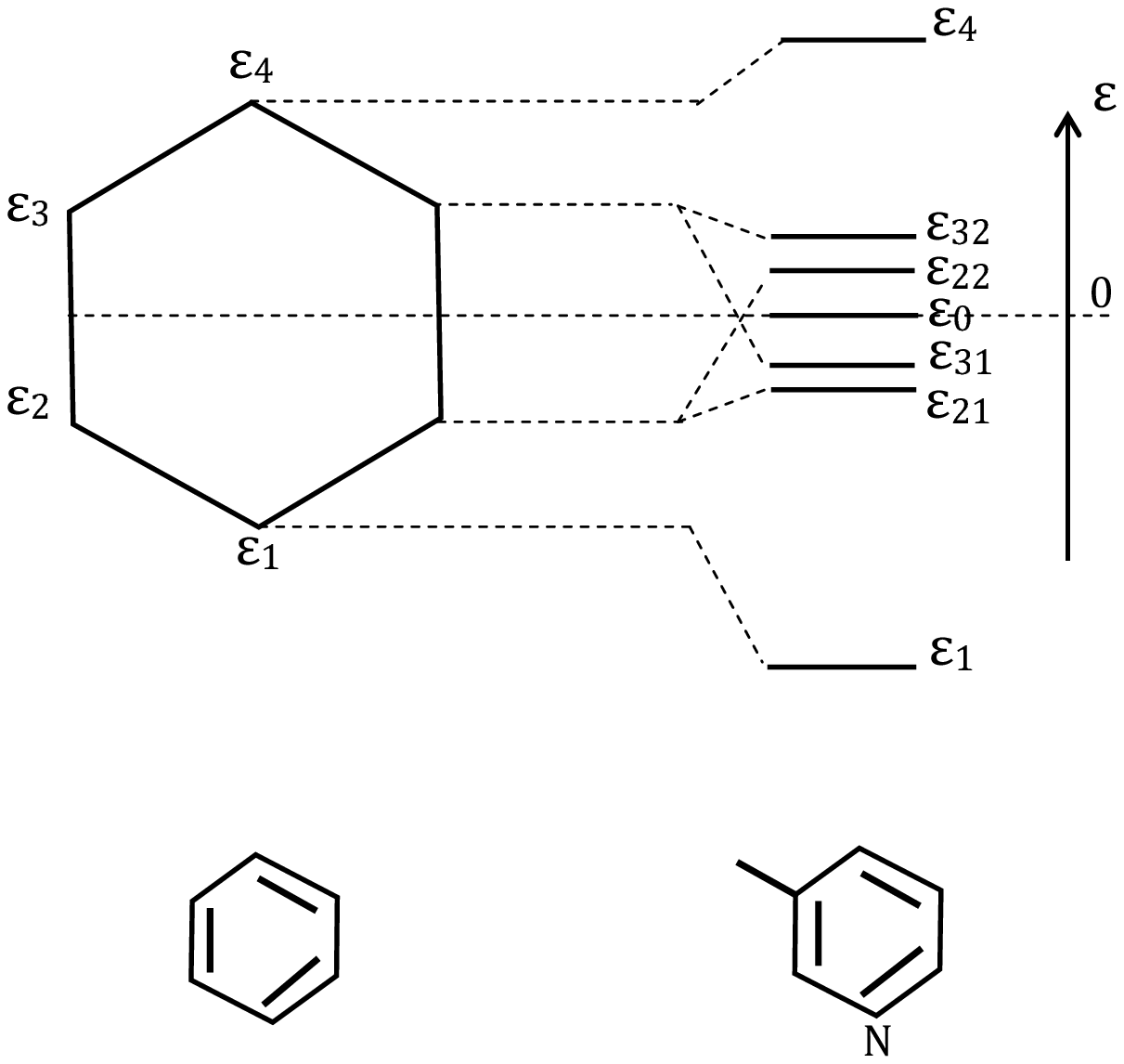}\protect\caption{The energy spectra of (left) benzene and (right) simplified cotinine
molecule. \label{fig:three}}
\end{figure}

In order to obtain $\left|\psi_{k}\right\rangle $ and $\varepsilon_{k}$
, we use the perturbation theory in quantum mechanics. We assume that
the unperturbed system is a benzene and an isolated methyl, i.e.

\begin{eqnarray}
H_{0} & = & \sum_{j=1}^{7}\alpha\left|j\right\rangle \left\langle j\right|+\sum_{j=1}^{6}\beta_{j,j+1}\left(\left|j\right\rangle \left\langle j+1\right|+\left|j+1\right\rangle \left\langle j\right|\right)\nonumber \\
 & = & \sum_{k=1}^{6}\varepsilon_{k}^{(0)}\left|k\right\rangle \left\langle k\right|+\alpha\left|7\right\rangle \left\langle 7\right|,
\end{eqnarray}
where
\begin{eqnarray}
\left|k\right\rangle  & = & \frac{1}{\sqrt{6}}\sum_{j=1}^{6}e^{-ikj}\left|j\right\rangle ,\\
\varepsilon_{k}^{(0)} & = & \alpha+2\beta\cos k
\end{eqnarray}
with momentum $k=m\pi/3$ and $m=-3,-2,-1,0,1,2$. For simplicity,
we use $\alpha=\alpha_{CC}$ and $\beta=\beta_{CC}$. The isolated
methyl contributes an isolated eigen state $\left|7\right\rangle $
with eigen energy $\alpha$. Then, the perturbation originates from
\begin{equation}
H^{'}=\mathcal{H}-H_{0}=\varDelta\alpha\left(a_{1}^{\dagger}a_{1}\right)+\varDelta\beta\left(a_{1}^{\dagger}a_{2}+a_{2}^{\dagger}a_{1}+a_{1}^{\dagger}a_{6}+a_{6}^{\dagger}a_{1}\right)+\beta\left(a_{3}^{\dagger}a_{7}+a_{7}^{\dagger}a_{3}\right),
\end{equation}
where $a_{j}^{\dagger}\left|0\right\rangle =\left|j\right\rangle $
and $a_{j}\left|j\right\rangle =\left|0\right\rangle $ with $\left|0\right\rangle $
being the vacuum state, $\Delta\alpha=\alpha_{N}-\alpha_{C}$ and
$\Delta\beta=\beta_{CN}-\beta_{CC}$. $a_{j}^{\dagger}$ is the creation
operator on $j$th site and $a_{j}$ is the annihilation operator.
Benzene has four energy levels $\varepsilon_{1},\varepsilon_{2},\varepsilon_{3},\varepsilon_{4}$,
which are labeled sequencially from the lowest eigen energy. Both
the degeneracies of $\varepsilon_{2}$ and $\varepsilon_{3}$ are
two, while $\varepsilon_{1}$ and $\varepsilon_{4}$ are non-degenerate.
Because one nitrogen atom substitutes for one carbon atom in the pyridyl,
the degeneracy is lifted. As a consequence, the simplified cotinine
molecule has seven energy levels (see Figure \ref{fig:three}), and
all of them are non-degenerate with the methyl giving the additional
$\varepsilon_{0}$. For the sake of simplicity, we further assume
$\alpha=0$.

According to the perturbation theory, the energy spectrum of the simplified
cotinine molecule reads
\begin{equation}
\begin{array}[t]{ccc}
\varepsilon_{1}^{(1)} & = & \frac{1}{6}\Delta\alpha+\frac{2}{3}\Delta\beta+\frac{2\beta}{\sqrt{6}},\\
\varepsilon_{21}^{(1)} & = & \frac{1}{6}\left[\Delta\alpha-2\Delta\beta+2\sqrt{6}\beta-\sqrt{\Delta\alpha^{2}+2\Delta\alpha\left(\Delta\beta+2\sqrt{6}\beta\right)+\left(\Delta\beta^{2}+4\sqrt{6}\Delta\beta\beta+24\beta^{2}\right)}\right],\\
\varepsilon_{22}^{(1)} & = & \frac{1}{6}\left[\Delta\alpha-2\Delta\beta+2\sqrt{6}\beta+\sqrt{\Delta\alpha^{2}+2\Delta\alpha\left(\Delta\beta+2\sqrt{6}\beta\right)+\left(\Delta\beta^{2}+4\sqrt{6}\Delta\beta\beta+24\beta^{2}\right)}\right],\\
\varepsilon_{31}^{(1)} & = & \frac{1}{6}\left[\Delta\alpha+2\Delta\beta-2\sqrt{6}\beta-\sqrt{\Delta\alpha^{2}-2\Delta\alpha\left(\Delta\beta+2\sqrt{6}\beta\right)+\left(\Delta\beta^{2}+4\sqrt{6}\Delta\beta\beta+24\beta^{2}\right)}\right],\\
\varepsilon_{32}^{(1)} & = & \frac{1}{6}\left[\Delta\alpha+2\Delta\beta-2\sqrt{6}\beta+\sqrt{\Delta\alpha^{2}+2\Delta\alpha\left(\Delta\beta+2\sqrt{6}\beta\right)+\left(\Delta\beta^{2}+4\sqrt{6}\Delta\beta\beta+24\beta^{2}\right)}\right],\\
\varepsilon_{4}^{(1)} & = & \frac{1}{6}\Delta\alpha-\frac{2}{3}\Delta\beta-\frac{2\beta}{\sqrt{6}},\\
\varepsilon_{\left|7\right\rangle }^{(1)} & = & 0.
\end{array}
\end{equation}
Following the degenerate perturbation theory, we can obtain the wave
function to the first order. Since their explicit expressions are
very complicated, we do not list them here.

\begin{figure}
\includegraphics[bb=50bp 400bp 595bp 800bp,clip,scale=0.5]{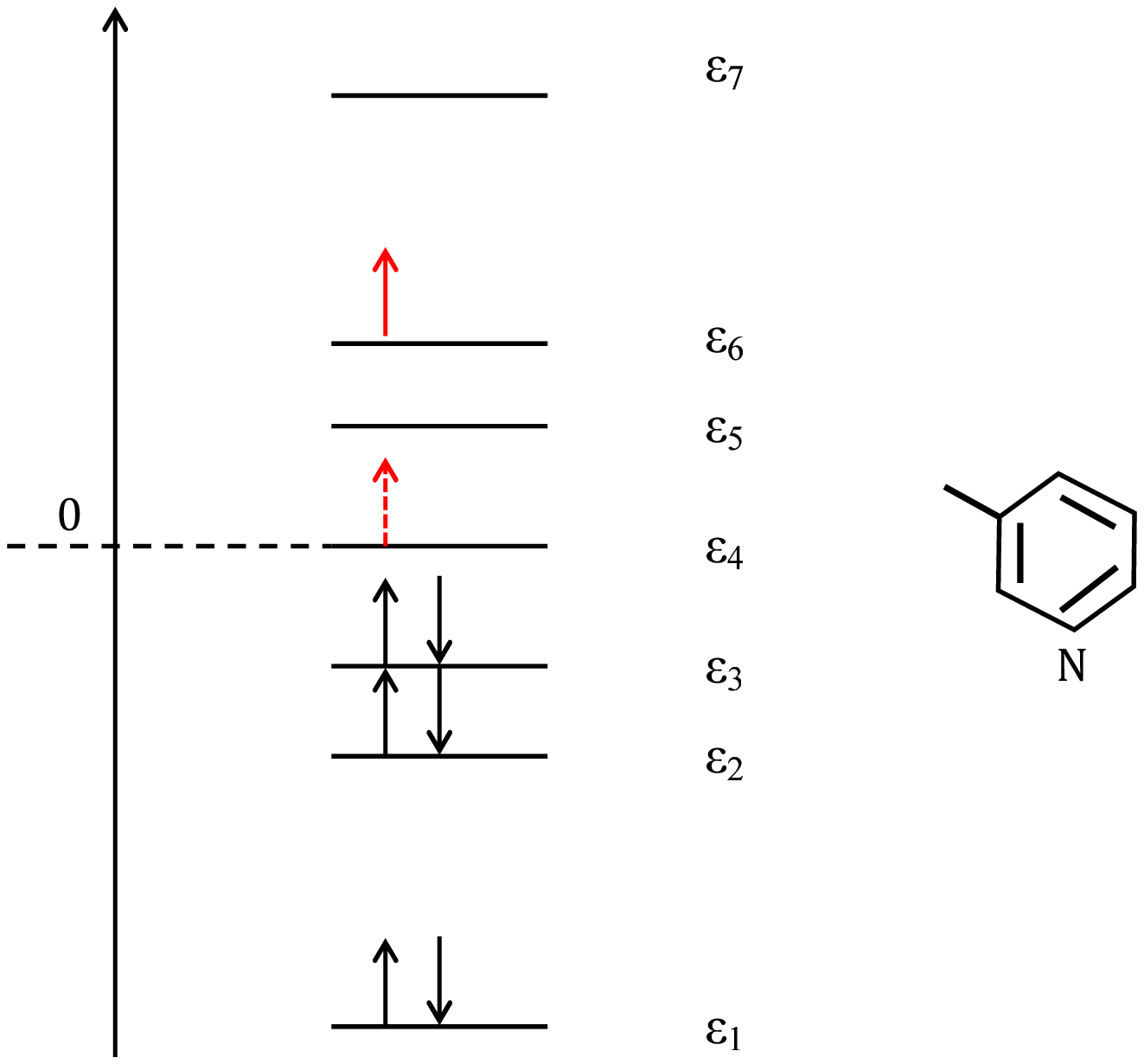}\protect\caption{The single-electron excitation of cotinine molecule (red solid up-arrow)
from the ground state (red dashed up-arrow). \label{fig:three-1}}
\end{figure}

The cotinine molecule has seven non-interacting $\pi$-electrons which
fill in seven energy levels. The energy levels are illustrated in
Figure~\ref{fig:three}. On account of the spin degree (see Figure
\ref{fig:three-1}), the ground state can be expressed in the second-quantization
form as

\begin{equation}
\left|\Psi_{0}\right\rangle =a_{1\uparrow}^{\dagger}a_{1\downarrow}^{\dagger}a_{2\uparrow}^{\dagger}a_{2\downarrow}^{\dagger}a_{3\uparrow}^{\dagger}a_{3\downarrow}^{\dagger}a_{4\uparrow}^{\dagger}\left|0\right\rangle ,
\end{equation}
and we use $E_{0}$ to represent the ground-state energy of the whole
cotinine system, i.e.
\begin{equation}
E_{0}=2(\varepsilon_{1}+\varepsilon_{2}+\varepsilon_{3})+\varepsilon_{4},
\end{equation}
where $a_{k\sigma}^{\dagger}$ is the creation operator of the orbital
$k$ with spin $\sigma$ ($\sigma=\uparrow,\downarrow$).

The system has twenty seven single-excitation states, for example,
the 2th and 11th excitation states are
\[
\begin{array}[t]{ccc}
\left|\Psi_{2}\right\rangle  & = & a_{1\uparrow}^{\dagger}a_{1\downarrow}^{\dagger}a_{2\uparrow}^{\dagger}a_{2\downarrow}^{\dagger}a_{3\uparrow}^{\dagger}a_{3\downarrow}^{\dagger}a_{6\uparrow}^{\dagger}\left|0\right\rangle ,\\
\left|\Psi_{11}\right\rangle  & = & a_{1\uparrow}^{\dagger}a_{1\downarrow}^{\dagger}a_{2\uparrow}^{\dagger}a_{2\downarrow}^{\dagger}a_{3\uparrow}^{\dagger}a_{7\downarrow}^{\dagger}a_{4\uparrow}^{\dagger}\left|0\right\rangle ,
\end{array}
\]
 with corresponding eigen energies
\[
\begin{array}[t]{ccc}
E_{2} & = & 2\left(\varepsilon_{1}+\varepsilon_{2}+\varepsilon_{3}\right)+\varepsilon_{6},\\
E_{11} & = & 2\left(\varepsilon_{1}+\varepsilon_{2}\right)+\varepsilon_{3}+\varepsilon_{4}+\varepsilon_{7},
\end{array}
\]
respectively. In the first case, the electron with energy $\varepsilon_{4}$
and spin up is excited to energy level $\varepsilon_{6}$. In the
second case, the electron with energy $\varepsilon_{3}$ and spin
down is excited to energy level $\varepsilon_{7}$. Here the flip
of electronic spin is not taken into consideration. To sum up, the
single-excitation states read
\begin{equation}
\left|\Psi_{n}\right\rangle =\left|\Psi_{p\sigma}^{q\sigma}\right\rangle =a_{q\sigma}^{\dagger}a_{p\sigma}\left|\Psi_{0}\right\rangle ,
\end{equation}
where $p=1,2,3,4$, $q=4,5,6,7$ and $\sigma=\uparrow,\downarrow$,
an the eigen energies are
\begin{equation}
E_{n}=E_{0}+\varepsilon_{q}-\varepsilon_{p}.
\end{equation}

In the subspaces spanned by the ground state and single-excitation
states, the Hamiltonian without electro-magnetic field reads

\begin{equation}
\mathcal{H}=\sum_{n=0}^{27}E_{n}\left|\Psi_{n}\right\rangle \left\langle \Psi_{n}\right|.
\end{equation}

\subsection*{Perturbation Theory in Electromagnetic Field}
When there is a time-dependent electromagnetic field applied on the
molecule, based on the dipole approximation, the total Hamiltonian
including the interaction between the molecule and the electromagnetic
field can be written as
\begin{eqnarray}
H & = & \mathcal{H}-\vec{\mu}\cdot\vec{E}(\vec{r},t)-\vec{m}\cdot\vec{B}(\vec{r},t)\nonumber \\
 & \simeq & \mathcal{H}-\vec{\mu}\cdot\vec{E}_{0}\cos(\vec{k}\cdot\vec{r}-\omega t)-\vec{m}\cdot\vec{B}_{0}\cos(\vec{k}\cdot\vec{r}-\omega t),
\end{eqnarray}
where$\mathcal{H}$ is the Hamiltonian without electromagnetic field,
$\vec{\mu}$ and $\vec{m}$ denote the electric and magnetic dipole
moments respectively. By assuming the spatial scale of the molecule
is much smaller than the wave length of the field $\vec{k}\cdot\vec{r}\simeq0$
(the coordinate is chosen as Figure \ref{fig:two}(c)), we have

\begin{equation}
H\simeq\mathcal{H}-\vec{\mu}\cdot\vec{E}_{0}\cos(\omega t)-\vec{m}\cdot\vec{B}_{0}\cos(\omega t).
\end{equation}

By a unitary transformation
\begin{equation}
U^{\dagger}=\exp\left(-i\omega\left|\Psi_{0}\right\rangle \left\langle \Psi_{0}\right|t\right),
\end{equation}
the Hamiltonian becomes time-independent
\begin{eqnarray}
H^{'} & = & U^{\dagger}HU-iU^{\dagger}\dot{U}\nonumber \\
 & \simeq & \sum_{n=1}^{27}E_{n}\left|\Psi_{n}\right\rangle \left\langle \Psi_{n}\right|+\left(E_{0}+\omega\right)\left|\Psi_{0}\right\rangle \left\langle \varPsi_{0}\right|+H^{\prime\prime},
\end{eqnarray}
where
\begin{eqnarray}
H^{\prime\prime} & = & -\frac{1}{2}\sum_{n=1}^{27}\left(\vec{\mu}_{n0}\cdot\vec{E}_{0}\left|\varPsi_{n}\right\rangle \left\langle \varPsi_{0}\right|+\vec{\mu}_{0n}\cdot\vec{E}_{0}\left|\varPsi_{0}\right\rangle \left\langle \varPsi_{n}\right|\right)\nonumber \\
 &  & -\frac{1}{2}\sum_{n=1}^{27}\left(\vec{m}_{n0}\cdot\vec{B}_{0}\left|\Psi_{n}\right\rangle \left\langle \Psi_{0}\right|+\vec{m}_{0n}\cdot\vec{B}_{0}\left|\Psi_{0}\right\rangle \left\langle \Psi_{n}\right|\right),\\
\vec{\mu}_{nn^{\prime}} & = & \left\langle \Psi_{n}\right|\vec{\mu}\left|\Psi_{n^{\prime}}\right\rangle ,\\
\vec{m}_{nn^{\prime}} & = & \left\langle \Psi_{n}\right|\vec{m}\left|\varPsi_{n^{\prime}}\right\rangle .
\end{eqnarray}
In other words, we change the system from the static frame into a
rotating frame. In the rotating frame, the state and operator become
$\left|\Psi^{\prime}\right\rangle =U^{\dagger}\left|\Psi\right\rangle ,$
and $A^{\prime}=U^{\dagger}AU$, respectively. Moreover, due to the
interaction with electromagnetic field,the molecular ground state
becomes
\begin{equation}
\left|\Psi_{0}^{\prime}\right\rangle =U^{\dagger}\left|\Psi_{0}\right\rangle =\left|\Psi_{0}\right\rangle +\sum_{n=1}^{27}\frac{\left\langle \Psi_{n}\right|H^{\prime\prime}\left|\Psi_{0}\right\rangle }{E_{0}+\omega-E_{n}}\left|\Psi_{n}\right\rangle ,
\end{equation}

\subsubsection*{Permittivity}

The electric dipole moment in the rotating frame reads
\begin{equation}
\vec{\mu}^{\prime}=U^{\dagger}\vec{\mu}U=\sum_{n=1}^{27}\left(\vec{\mu}_{n0}e^{i\omega t}\left|\Psi_{n}\right\rangle \left\langle \Psi_{0}\right|+\vec{\mu}_{0n}e^{-i\omega t}\left|\varPsi_{0}\right\rangle \left\langle \Psi_{n}\right|\right).
\end{equation}
 For the ground state, the expectation value for the dipole operator
in the rotating frame is
\begin{equation}
\left\langle \Psi_{0}^{'}\right|\vec{\mu}^{\prime}\left|\Psi_{0}^{'}\right\rangle =-\textrm{Re}\sum_{n=1}^{27}\frac{\vec{\mu}_{n0}\cdot\vec{E}_{0}}{E_{0}+\omega-E_{n}}\vec{\mu}_{0n}e^{-i\omega t}.
\end{equation}
 In the electromagnetic field, the electric displacement field in
a volume $V$ with $N$ identical molecules

\begin{eqnarray}
\vec{D} & = & \varepsilon\vec{E}_{0}\nonumber \\
 & = & \varepsilon_{0}\varepsilon_{r}\vec{E}_{0}\nonumber \\
 & = & \varepsilon_{0}\vec{E}_{0}+\frac{\vec{P}}{V}
\end{eqnarray}
reads
\begin{equation}
\vec{D}=\varepsilon_{0}\vec{E}_{0}-\sum_{s=1}^{N}\sum_{n=1}^{27}\frac{\left[\vec{\mu}_{n0}(s)\cdot\vec{E}_{0}\right]\vec{\mu}_{0n}(s)}{V\left(E_{0}+\omega-E_{n}\right)}.
\end{equation}
Thus, the total permittivity in different direction is
\begin{equation}
\varepsilon_{ij}=\varepsilon_{0}\delta_{ij}-\sum_{s=1}^{N}\sum_{n=1}^{27}\frac{\mu_{0n}^{(i)}(s)\mu_{n0}^{(j)}(s)}{\left(E_{0}+\omega-E_{n}\right)V},\textrm{ for }i,j=x,y,z
\end{equation}
The relative dielectric constant of the system, i.e. the permittivity,
gives
\begin{equation}
\varepsilon_{ij}^{r}\equiv\delta_{ij}-\sum_{s=1}^{N}\sum_{n=1}^{27}\frac{\vec{\mu}_{0n}(s)\cdot\hat{e}_{i}\vec{\mu}_{n0}(s)\cdot\hat{e}_{j}}{\varepsilon_{0}V\left(E_{0}+\omega-E_{n}\right)},\textrm{ for }i,j=x,y,z,\label{eq:ep}
\end{equation}
where $\hat{e}_{i}$ is the unit vector of the lab coordinate system.

If we choose the symmetric center of benzene as the origin of coordinate
(see Figure \ref{fig:two}(c)), the electric dipole moment reads
\begin{equation}
\vec{\mu}=-\sum_{j=1}^{7}e\vec{r}_{j},
\end{equation}
where $\vec{r}_{j}$ is the vector of $j$th electron and $-e$ is
the electric charge of electrons. Because $\vec{r}_{j}$'s are single-electron
operators, the matrix elements of electric dipole operators are given
by
\begin{eqnarray}
\vec{\mu}_{0n}=\left\langle \Psi_{0}\right|\vec{\mu}\left|\Psi_{n}\right\rangle  & = & -e\left\langle \Psi_{0}\right|\vec{r}\left|\Psi_{n}\right\rangle =-e\left\langle \psi_{p}\right|\vec{r}\left|\psi_{q}\right\rangle =-e\vec{r}_{pq},
\end{eqnarray}
where $\vec{r}_{pq}$ is the overlap of $\vec{r}$ between two single-electron
wave functions, i.e.
\begin{equation}
\left\langle \psi_{p}\right|\vec{r}\left|\psi_{q}\right\rangle =\sum_{j=1}^{7}C_{pj}^{*}C_{qj}\vec{r}_{j}.
\end{equation}

\subsubsection*{Permeability}

To account for the magnetic response of cotinine molecule, we start
from the Heisenberg equations of motion,
\begin{eqnarray}
p^{x} & = & m_{e}\dot{r}^{x}=im_{e}\left[\mathcal{H},r^{x}\right],\\
p^{y} & = & m_{e}\dot{r}^{y}=im_{e}\left[\mathcal{H},r^{y}\right],
\end{eqnarray}
where we assume $\hbar=1$ and $m_{e}$ is the mass of electrons of
cotinine molecule.

The magnetic dipole moment is related to the angular momentum of the
system. The angular momentum operators read
\begin{eqnarray}
L_{x} & = & r^{y}p^{z}-r^{z}p^{y}=0,\\
L_{y} & = & r^{z}p^{x}-r^{x}p^{z}=0,\\
L_{z} & = & r^{x}p^{y}-r^{y}p^{x}\nonumber \\
 & = & \frac{1}{2}\left(r^{x}p^{y}+p^{y}r^{x}-r^{y}p^{x}-p^{x}r^{y}\right)\nonumber \\
 & = & im_{e}\left(r^{x}\mathcal{\mathcal{H}}r^{y}-r^{y}\mathcal{H}r^{x}\right).
\end{eqnarray}
Obiviously, only the response in $z$ direction is present as all
atoms in the cotinine molecule are restricted in the $xy$ plane (see
Figure \ref{fig:two}(c)). Therefore, the magnetic dipole moment is
\begin{eqnarray}
\vec{m} & = & \frac{-e}{2m_{e}}\vec{L}\nonumber \\
 & = & \frac{-e}{2m_{e}}L_{z}\hat{e}_{z}\nonumber \\
 & = & \frac{-ie}{2}\left(r^{x}\mathcal{H}r^{y}-r^{y}\mathcal{H}r^{x}\right)\hat{e}_{z}\nonumber \\
 & = & \frac{-ie}{2}\sum_{k,k^{\prime}}\sum_{n}E_{n}\left(r_{kn}^{x}r_{nk^{\prime}}^{y}-r_{kn}^{y}r_{nk^{\prime}}^{x}\right)\left|\Psi_{k}\right\rangle \left\langle \Psi_{k^{'}}\right|\hat{e}_{z},
\end{eqnarray}
where
\begin{equation}
r_{kk^{\prime}}^{\alpha}=\left\langle \psi_{k}\right|r^{\alpha}\left|\psi_{k^{\prime}}\right\rangle .
\end{equation}

Similar to the electric response, the expectation value for the magnetic
dipole operator in the rotating frame is
\begin{equation}
\left\langle \Psi_{0}^{'}\right|\vec{m}^{\prime}\left|\Psi_{0}^{'}\right\rangle =-\textrm{Re}\sum_{n=1}^{27}\frac{\vec{m}_{n0}\cdot\vec{B}_{0}}{E_{0}+\omega-E_{n}}\vec{m}_{0n}e^{-i\omega t}.
\end{equation}
The magnetic induction in a volume $V$ with $N$ identical molecules

\begin{eqnarray}
\vec{B} & = & \mu\vec{H}_{0}\nonumber \\
 & = & \mu_{0}\mu_{r}\vec{H}_{0}\nonumber \\
 & = & \mu_{0}\vec{H}_{0}+\mu_{0}\frac{\vec{M}}{V}
\end{eqnarray}
is explicitly given by
\begin{equation}
\vec{B}=\mu_{0}\vec{H}_{0}-\sum_{s=1}^{N}\sum_{n=1}^{27}\frac{\mu_{0}\vec{m}_{n0}(s)\cdot\vec{B}_{0}}{V\left(E_{0}+\omega-E_{n}\right)}\vec{m}_{0n}(s).
\end{equation}
Notice that $\mu$ is the permeability of medium, different from the
electric dipole moment $\vec{\mu}$ above. And $\vec{H}_{0}$ is magnetic
field intensity, not the Hamiltonian without electromagnetic field
$H_{0}$.

The relative permeability of cotinine system is simplified as
\begin{equation}
\mu_{ij}^{r}\equiv\delta_{ij}-\mu_{0}\sum_{s=1}^{N}\sum_{n=1}^{27}\frac{\vec{m}_{0n}(s)\cdot\hat{e}_{i}\vec{m}_{n0}(s)\cdot\hat{e}_{j}}{V\left(E_{0}+\omega-E_{n}\right)},\textrm{ for }i,j=x,y,z.\label{eq:mu}
\end{equation}

\subsubsection*{Analysis}

Equations (\ref{eq:ep}) and (\ref{eq:mu}) present the analytical
results of the permittivity and permeability of cotinine molecule
in electromagnetic field. According to the expressions of these two
quantities, they can be negative simultaneously when the second parts
of the expressions greater than unity. In order to fulfill this requirement,
the denominators of the second part should be small enough. In other
words, $E_{0}+\omega-E_{n}$ needs to be much smaller than numerator
which means $\omega\approx E_{n}-E_{0}$. For a given initial energy
of the electron before transition $E_{0}$, we can observe simultaneous
negative permittivity and permeability of cotinine molecules in electromagnetic
field when the driving frequency $\omega$ is tuned approximately
equal to the transition frequency $E_{n}-E_{0}$ .

\subsection*{Numerical Simulation of Permittivity and Permeability\label{sec:result}}

\begin{figure}
\includegraphics[scale=0.4]{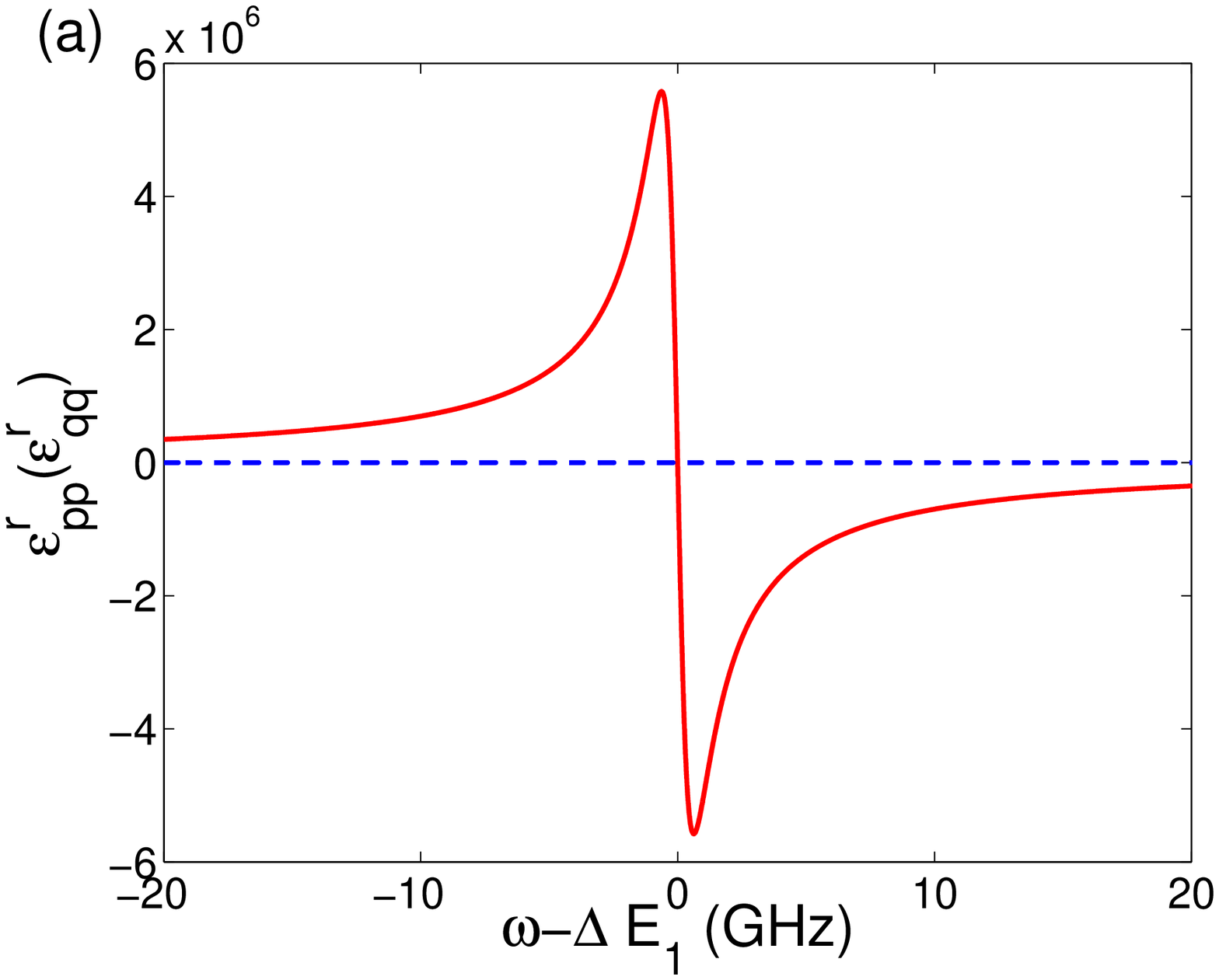}\includegraphics[scale=0.4]{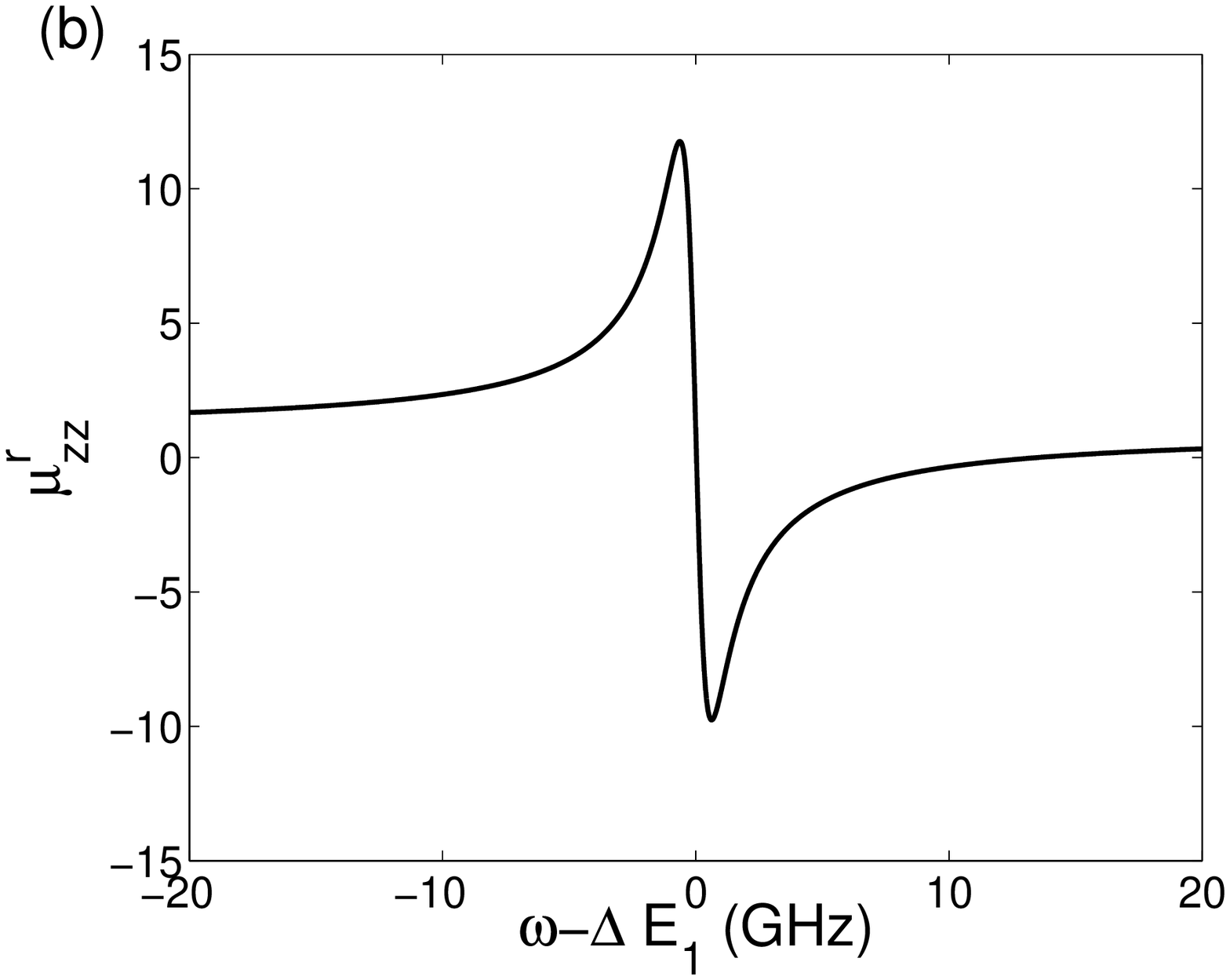}\protect\caption{The numerical results of (a) permittivity and (b) permeability of
cotinine molecules vs the light frequency $\omega$. The permittivity
$\varepsilon_{oo}^{r}$ (red solid line) along one main axis in the
$xy$ plane is negative near the resonance frequency, while $\varepsilon_{pp}^{r}$
(blue dashed line) along the other main axis is always constant. In
the magnetic response, the permeability along $z$ direction $\mu_{zz}^{r}$
is shown. \label{fig:re1}}
\end{figure}

In above section, the analytical derivation suggests
that permittivity and permeability of cotinine molecules might be
negative simultaneously in certain frequency regime. Here we show
and analyze the numerical result. In the simplified model, the cotinine
molecule is simplified into a pyridine and a methyl, c.f. Figure \ref{fig:two}
(b). The simplified cotinine model is a two dimensional model. Thus,
we only need to analyze the electromagnetic responses of the molecules
in $x$ and $y$ directions. Figure \ref{fig:re1} shows the numerical
simulation of relative dielectric constants in the $xy$ plane and
relative magnetic permittivity in $z$ direction of the system. Here
we assume the site energies $\alpha_{C}=0$, $\alpha_{N}=0.5\beta_{CC}$,
and the coupling strengths $\beta_{CC}=-3.6$eV, $\beta_{CN}=0.8\beta_{CC}$
\cite{x12}. The excited-state life time $\tau=10$ns is within the
range of experimentally observation, e.g. $90\mu$s \cite{Tokuji09}.
For a transition to the first excited state, e.g. a spin up electron
is excited from $\varepsilon_{3}$ to $\varepsilon_{4}$, the contributions
from transition dipoles $\mu_{01}$ and $m_{01}$ are much larger
than others i.e., $\omega\sim\Delta E_{1}=E_{1}-E_{0}$. In Figure
\ref{fig:re1}(a), both relative dielectric contants in the two main
axes $\varepsilon_{pp}^{r}$ and $\varepsilon_{qq}^{r}$ are different
from unity in the vaccum case, as the presence of nitrogen atom breaks
down the reflection symmetry along the axis connecting site 3(7) and
the origin. Furthermore, Figure \ref{fig:re1} clearly shows the negative
permittivity and permeability at the same time. This result suggests
that cotinine molecules can be detected by negative refraction.

\section*{Discussion}

In this paper we research the optical properties of drug metabolites
in latent fingermarks. All of these drug metabolites have a structure
in common, i.e. SRR which could realize negative refraction. And negative
refraction makes the optical properties of latent fingermark
quite different between drug addicts and non-drug users which can
be used to distinguish them. The latent fingermarks of these two kinds
of donors are observed to be inverse in light field. The method is printing the donor's fingermarks on the transparent media and observing them in the light transmission direction on the opposite side with respect to the side for the normal refraction. Because of negative refraction, the fingermarks of drug addicts are easily observable. 
Without loss of generality, we take cotinine as an example to calculate electromagnetic
response of metabolites in latent fingermarks of smokers. According
to our analytic derivation and numerical simulation, we demonstrate
the presence of negative refraction in cotinine molecules. The advantage
of this method is that it is physical and non-damaged. Our method
is suitable for all drug metabolites which have the SRR structure.
And this method can be conveniently applied to distinguish drug addicts
and non-drug users too. For example, except for cotinine, benzoylecgonine
and morphine can also be detected using our method.


\section*{Acknowledgements}

The research was supported by Open Research Fund Program of the State
Key Laboratory of Low-Dimensional Quantum Physics, Tsinghua University
Grant No.~KF201502.

\section*{Author contributions statement}

Y.S. wrote the main manuscript text  and did the calculations. Y.S. and Q.A. designed the project and reviewed the manuscript.

\end{document}